# Correlation Between Gravitational and Inertial Mass: Theory and Experimental Test.

## Fran De Aquino
Maranhao State University,
Physics Department,
65058-970 S.Luis/MA, Brazil.

**The physical property of mass has two distinct aspects, gravitational mass and inertial mass. The weight of a particle depends on its gravitational mass. According to the weak form of the equivalence principle, the gravitational and inertial masses are equivalent. But, we show here that they are correlated by a dimensionless factor, which can be different of one . The factor depends on the electromagnetic energy absorbed or emitted by the particle and of the index of refraction of the medium around the particle. This theoretical correlation has been experimentally proven by means of a very simple apparatus, presented here.**

The *gravitational mass*, $m_g$ , produces and responds to gravitational fields. It supplies the mass factors in Newton's famous inverse-square law of Gravitation $\left(F_{12} = Gm_{g1}m_{g2}/r_{12}^2\right)$. Inertial mass $m_i$ is the mass factor in Newton's 2nd Law of Motion $(F = m_i a)$.

Several experiments[1-6], have been carried out since Newton to try to establish a correlation between gravitational mass and inertial mass.

Recently J.F.Donoghue and B.R. Holstein[7] have shown that the renormalized mass for temperature $T = 0$ is expressed by $m_r = m + \delta m_0$ where $\delta m_0$ is the *temperature-independent mass shift*. In addition, for $T > 0$, mass renormalization leads to the following expressions for inertial and gravitational masses, respectively: $m_i = m + \delta m_0 + \delta m_\beta$ ; $m_g = m + \delta m_0 - \delta m_\beta$, where $\delta m_\beta$ is the *temperature-dependent mass shift* given by $\delta m_\beta = \pi\alpha T^2/3m_i$ .

This means that a particle's gravitational mass decreases with the increasing temperature and that only in absolute zero $(T = 0K)$ are gravitational mass and inertial mass equivalent.

The expression of $\delta m_\beta$ obtained by Donoghue and Holstein refers solely to thermal radiation. But,

$\delta m_\beta$ represents also *the inertial mass shift* . It can be seen by repeating the renormalization of the external electromagnetic vertex at finite temperature. On the other hand, it is easy to see that *the inertial mass shift* is related to *inertial Hamiltonian shift* $\delta H$ . Thus we can obtain the *general expression* of $\delta m_\beta$ by means of the *inertial Hamiltonian shift* $\delta H$ ,i.e.,

$$\delta m_\beta = \frac{\delta H}{c^2} = \frac{c\sqrt{\delta p^2 + m_i^2 c^2} - m_i c^2}{c^2} =$$

$$= m_i\sqrt{1 + \left(\frac{\delta p}{m_i c}\right)^2} - m_i \qquad [1]$$

where $\delta p$ is the correspondent particle's *momentum* shift.

Consequently, the general expression of correlation between gravitational and inertial mass can be write in the following form

$$m_g = m_i - 2\delta m_\beta = \left\{1 - 2\left[\sqrt{1 + \left(\frac{\delta p}{m_i c}\right)^2} - 1\right]\right\}m_i \quad [2]$$

We can look on this change in *momentum* as due to the electromagnetic energy *absorbed* or *emitted* by the particle ( absorbed or emitted radiation by the particle and/or *Lorentz's force* upon *charged* particle due to electromagnetic field).

In the case of radiation, according to Quantum Mechanics, we can write

$$\delta p = N\hbar k = N\hbar\omega / (\omega / k) = U / (dz / dt) =$$
$$= U / v \qquad [3]$$

Where $U$ is *the electromagnetic energy absorbed or emitted by the particle* and $v$ is the velocity of the electromagnetic waves, which can be write as follows

$$v = \frac{c}{\sqrt{\frac{\varepsilon_r \mu_r}{2}\left(\sqrt{1+(\sigma/\omega\varepsilon)^2}+1\right)}} \qquad [4]$$

$\varepsilon$ , $\mu$ and $\sigma$, are the electromagnetic characteristics of the medium in which the incident (or emitted) radiation is propagating ( $\varepsilon = \varepsilon_r \varepsilon_0$ where $\varepsilon_r$ is the *relative electric permittivity* and $\varepsilon_0 = 8.854 \times 10^{-12} F / m$ ; $\mu = \mu_r \mu_0$ where $\mu_r$ is the *relative magnetic permeability* and $\mu_0 = 4\pi \times 10^{-7} H / m$ ). For an *atom* inside a body , the incident(or emitted) radiation on this atom will be propagating inside the body , and consequently , $\sigma = \sigma_{body}$ , $\varepsilon = \varepsilon_{body}$, $\mu = \mu_{body}$.

From the Eq.(3) follows that

$$\delta p = \frac{U}{v} = \frac{U}{c}\left(\frac{c}{v}\right) = \frac{U}{c}n_r \qquad [5]$$

where $n_r$ is the *index of refraction*, given by

$$n_r = \frac{c}{v} = \sqrt{\frac{\varepsilon_r \mu_r}{2}\left(\sqrt{1+(\sigma/\omega\varepsilon)^2}+1\right)} \qquad [6]$$

$c$ is the speed in a vacuum and $v$ is the speed in the medium.

By the substitution of Eq.(5) into Eq.(2), we obtain

$$m_g = \left\{1 - 2\left[\sqrt{1+\left(\frac{U}{m_i c^2}n_r\right)^2}-1\right]\right\}m_i \qquad [7]$$

Recently, L.V. Hau *et al.*,[8] succeeded in reducing the speed of light to 17 m/s by optically inducing a quantum interference in a *Bose-Einstein condensate*. This means an enormous index of refraction ( $n_r \approx 10^7$ ) at ~$10^{14}$Hz.

Light can be substantially slowed down or frozen completely [9]. If t he speed of light is reduced to <0.1m/s, the Eq.(7) tell us that the gravitational masses of the *atoms* of the Bose-Einstein condensate become *negative*.

If the absorbed (or emitted) radiation is *monochromatic* and has frequency $f$ , we can put $U = nhf$ in Equation(7), where $n$ is the number of incident (or radiated) photons on the particle of mass $m_i$. Thus we obtain

$$m_g = m_i - 2\left\{\sqrt{1+\left\{\frac{nhf}{m_i c^2}n_r\right\}^2}-1\right\}m \qquad [8]$$

In that case, according to the *Statistical Mechanics*, the calculation of $n$ can be made based on the well-known method of *Distribution Probability* . If all the particles inside the body have the same mass $m_i$, the result is

$$n = \frac{N}{S}a \qquad [9]$$

where $S$ is the average density of absorbed (or emitted) photons on the body; $a$ is the area of the surface of a particle of mass $m_i$ from the body.

Obviously the power $P$ of the absorbed radiation must be $P = Nhf / \Delta t = Nhf^2$, thus we can write $N = P / hf^2$ . Substitution of $N$ into Eq.(9) gives

$$n = \frac{a}{hf^2}\left(\frac{P}{S}\right) = \frac{a}{hf^2}D \qquad [10]$$

where $D$ is the *power density* of the incident( or emitted) radiation. Thus Eq.(8) can be rewritten in the following form:

$$m_g = m_i - 2\left\{\sqrt{1+\left\{\frac{aD}{m_i cvf}\right\}^2}-1\right\}m \qquad [11]$$

For $\sigma >> \omega\varepsilon$ Eq.(4) reduces to

$$v = \sqrt{\frac{4\pi f}{\mu\sigma}} \qquad [12]$$

By substitution of Eq.(12) into Eq.(11) we obtain

$$m_g = m_i - 2\left\{\sqrt{1 + \left\{\frac{aD}{m_i c}\sqrt{\frac{\mu\sigma}{4\pi f^3}}\right\}^2} - 1\right\}m_i \quad [13]$$

This equation shows clearly that, *atoms* (or *molecules*) can have their *gravitational masses* strongly reduced by means of Extra-Low Frequency (ELF) radiation.

We have built an apparatus to produce ELF radiation ( transmitter and antenna) and to check the effects of this radiation upon the gravitational mass of a material surrounding the antenna ( see Fig. 1).

The antenna is a *half-wave dipole*, *encapsulated by a iron sphere* (purified iron, 99.95% Fe; $\mu_i = 5{,}000\mu_0$; $\sigma_i = 1.03 \times 10^7 S/m$ ).

The *radiation resistance* of the antenna for a frequency $\omega = 2\pi f$, can be written as follows [10]

$$R_r = \frac{\omega\mu_i\beta_i}{6\pi}\Delta z^2 \quad [14]$$

where $\Delta z$ is the length of the dipole and

$$\beta_i = \omega\sqrt{\frac{\varepsilon_i\mu_i}{2}\left(\sqrt{1 + (\sigma_i/\omega\varepsilon_i)^2} + 1\right)} =$$
$$= \frac{\omega}{c}\sqrt{\frac{\varepsilon_{ri}\mu_{ri}}{2}\left(\sqrt{1 + (\sigma_i/\omega\varepsilon_i)^2} + 1\right)} =$$
$$= \frac{\omega}{c}(n_r) = \frac{\omega}{c}\left(\frac{c}{v_i}\right) = \frac{\omega}{v_i} \quad [15]$$

where $v_i$ is the velocity of the radiation through the iron.

Substituting (15) into (14) gives

$$R_r = \frac{2\pi}{3}\left(\frac{\mu_i}{v_i}\right)(\Delta z f)^2 \quad [16]$$

Note that when the medium surrounding the dipole is *air* and $\omega >> \sigma/\varepsilon$, $\beta \cong \omega\sqrt{\varepsilon_0\mu_0}$, $v \cong c$ and $R_r$ reduces to the well-know expression $R_r \cong (\Delta z\omega)^2/6\pi\varepsilon_0 c^3$.

Here, due to $\sigma_i >> \omega\varepsilon_i$, $v_i$ is given by the Eq.(12). Then Eq.(16) can be rewritten in the following form

$$R_r = (\Delta z)^2\sqrt{\left(\frac{\pi}{9}\right)\sigma_i\mu_i^3 f^3} \quad [17]$$

The *ohmic resistance* of the dipole is [11]

$$R_{ohmic} \cong \frac{\Delta z}{2\pi r_0}R_S \quad [18]$$

where $r_0$ is the radius of the cross section of the dipole, and $R_S$ is the *surface resistance* ,

$$R_S = \sqrt{\frac{\omega\mu_{dipole}}{2\sigma_{dipole}}} \quad [19]$$

Thus,

$$R_{ohmic} \cong \frac{\Delta z}{r_0}\sqrt{\frac{\mu_{dipole} f}{4\pi\sigma_{dipole}}} \quad [20]$$

Where $\mu_{dipole} = \mu_{copper} \cong \mu_0$ and $\sigma_{dipole} = \sigma_{copper} = 5.8 \times 10^7 S/m$ .

Let us now consider the apparatus ( System H ) presented in Fig.1.

The *radiated power* for an *effective (rms)* current $I$ is then $P = R_r I^2$ and consequently

$$D = \frac{P}{S} = \frac{(\Delta z I)^2}{S}\sqrt{\left(\frac{\pi}{9}\right)\sigma_i\mu_i^3 f^3} \quad [21]$$

where $S$ is the *effective* area. It can be easily shown that $S$ is the outer area of the iron sphere, i.e., $S = 4\pi r_{outer}^2 = 0.19 m^2$ .

The iron surrounding the dipole increases its inductance $L$. However, for series RLC circuit the *resonance frequency* is $f_r = 1/2\pi\sqrt{LC}$ , then when $f = f_r$ ,

$$X_L - X_C = 2\pi f_r L - \frac{1}{2\pi f_r C} = \sqrt{\frac{L}{C}} - \sqrt{\frac{L}{C}} = 0.$$

Consequently, the impedance of the antenna, $Z_{ant}$, becomes *purely resistive*, i.e.,

$$Z_{ant} = \sqrt{R_{ant}^2 + (X_L - X_C)^2} = R_{ant} = R_r + R_{ohmic}.$$

For $f = f_r = 9.9 mHz$ the length of the dipole is

$$\Delta z = \lambda/2 = v/2f = \sqrt{\pi/\mu_i\sigma_i f} = 0.070 m = 70 mm.$$

Consequently, the *radiation resistance* $R_r$, according to Eq.(17), is $R_r = 4.56\mu\Omega$ and the *ohmic resistance*, for $r_0 = 13mm$, according to Eq.(20), is $R_{ohmic} \cong 0.02\mu\Omega$. Thus, $Z_{ant} = R_r + R_{ohmic} = 4.58\mu\Omega$ and the *efficiency* of the antenna is $e = R_r/R_r + R_{ohmic} = 99.56$ (99.56%).

The radiation of frequency $f = 9.9mHz$ is totally absorbed by the iron along a critical thickness $\delta = 5z = 5/\sqrt{\pi f\mu_i\sigma_i} \cong 0.11m = 110mm$. Therefore, from the Fig.1 we conclude that the iron sphere will absorb practically all radiation emitted from the dipole. Indeed, the sphere has been designed with this purpose, and in such a manner that all their atoms should be reached by the radiation. In this way, the radiation outside of the sphere is practically negligible.

When the ELF radiation strikes the iron atoms their gravitational masses, $m_{gi}$, are changed and, according to Eq.(13), become

$$m_{gi} = m_i - 2\left\{\sqrt{1 + \frac{\mu_i\sigma_i}{4\pi c^2 f^3}\left(\frac{a_i}{m_i}\right)^2 D^2} - 1\right\}m_i \quad [22]$$

Substitution of (21) into (22) yields

$$m_{gi} = m_i - 2\left\{\sqrt{1 + \left(\frac{\mu_i^2\sigma_i}{6cS}\right)\left(\frac{a_i}{m_i}\right)^2(\Delta I)^4} - 1\right\}m_i \quad [23]$$

Note that the equation above doesn't depends on $f$.

Thus, assuming that the radius of the iron atom is $r_{iron} = 1.40\times10^{-10}m$; $a_{iron} = 4\pi r_{iron}^2 = 2.46\times10^{-19}m^2$ and $m_{iron} = 55.85(1.66\times10^{-27}kg) = 9.27\times10^{-26}kg$ then the Eq.(23) can be rewritten as follows

$$m_{gi} = m_i - 2\left\{\sqrt{1 + 2.38\times10^{-4}I^4} - 1\right\}m_i \quad [24]$$

The equation above shows that the gravitational masses of the iron atoms can be nullified for $I \cong 8.51A$.

Above this critical current, $m_{gi}$ becomes *negative*.

The Table 1 presents the experimental results obtained from the System H for the gravitational mass of the *iron sphere*, $m_{g(iron\ sphere)}$, as a function of the current $I$, for $m_{iron\ sphere} = 60.50kg$ ( *inertial mass* of the iron sphere ). The values for $m_{g(iron\ sphere)}$, calculated by means of Eq.(24), are on that Table to be compared with those supplied by the experiment.

# APPENDIX: A Simple Derivation of the Correlation between Gravitational and Inertial Mass.

In order to obtain the general expression of correlation between $m_g$ and $m_i$, we will start with the definition of *inertial* Hamiltonian, $H_i$, and *gravitational* Hamiltonian, $H_g$, i.e.,

$$H_i = c\sqrt{p_i^2 + m_i^2 c^2} + Q\varphi \qquad [1]$$

$$H_g = c\sqrt{p_g^2 + m_g^2 c^2} + Q\varphi \qquad [2]$$

where $m_i$ and $m_g$ are respectively, the inertial and gravitational *masses at rest*; $p_i$ is the *inertial momentum* and $p_g$ the *gravitational momentum*; $Q$ is the electric charge and $\varphi$ is an electromagnetic potential.

A *momentum shift*, $\delta p$, on the particle, produces an *inertial Hamiltonian shift*, $\delta H$, given by

$$\delta H = \sqrt{(p_i + \delta p)^2 c^2 + m_i^2 c^4} - \sqrt{p_i^2 c^2 + m_i^2 c^4} \quad [3]$$

Fundamentally $\delta p$ is related to *absorption* or *emission* of energy.

In the general case of *absorption* and posterior *emission*, in which the particle acquires a $\delta p$ at the absorption and another $\delta p$ at the emission, the *total inertial* Hamiltonian *shift* is

$$\delta H = 2\left(\sqrt{(p_i + \delta p)^2 c^2 + m_i^2 c^4} - \sqrt{p_i^2 c^2 + m_i^2 c^4}\right) \quad [4]$$

Note that $\delta H$ is always *positive*.

We now may define the correlation between $H_i$ and $H_g$ as follows

$$H_i = H_g + \delta H \qquad [5]$$

If $\delta H = 0$, $H_i = H_g$, i.e., $m_g = m_i$.

In addition from the Eqs.[1] and [2], we can write:

$$H_i - H_g = \sqrt{p_i^2 c^2 + m_i^2 c^4} - \sqrt{p_g^2 c^2 + m_g^2 c^4} \quad [6]$$

For a particle at rest, $V = 0$;

$p_i = p_g = 0$. Consequently, Eqs.[4] and [6] reduces to

$$\delta H = 2\left(\sqrt{\delta p^2 c^2 + m_i^2 c^4} - m_i c^2\right) \qquad [7]$$

and

$$H_i - H_g = (m_i - m_g)c^2 \qquad [8]$$

Substitution of Eqs[7] and [8] into Eq.[5] yields

$$(m_i - m_g)c^2 = 2\left(\sqrt{\delta p^2 c^2 + m_i^2 c^4} - m_i c^2\right)$$

From this equation we obtain

$$m_g = m_i - 2\left[\sqrt{1 + \left(\frac{\delta p}{m_i c}\right)^2} - 1\right]m_i \qquad [9]$$

This is the general expression of correlation between gravitational and inertial mass.

Note that the term inside the square bracket is always positive. Thus, except for *anti-matter* ($m_i < 0$), the second term on the right hand side of Eq.[9] *is always negative*.

In particular, we can look on the *momentum shift* ($\delta p$) as due to absorption or emission of *electromagnetic energy* by the particle ( by means of *radiation* and/or by means of *Lorentz's force* upon the *charge* of the particle).

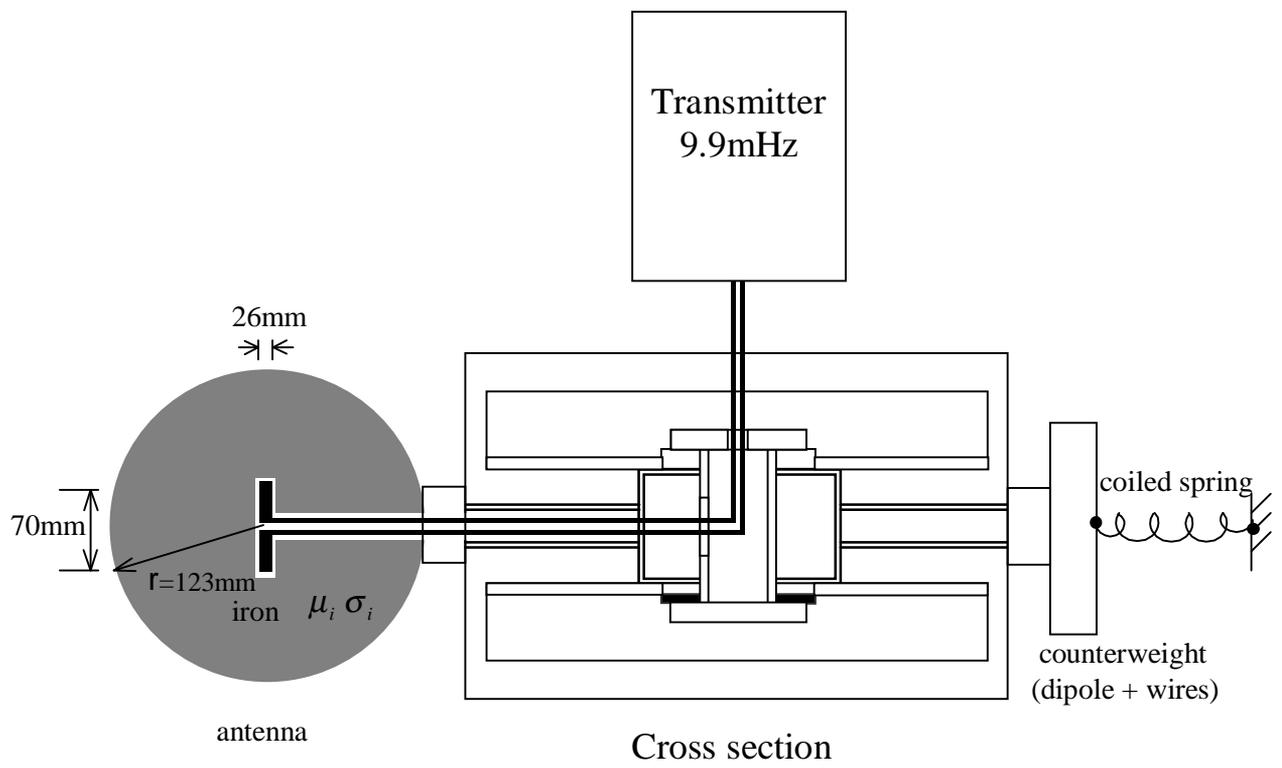

Cross section

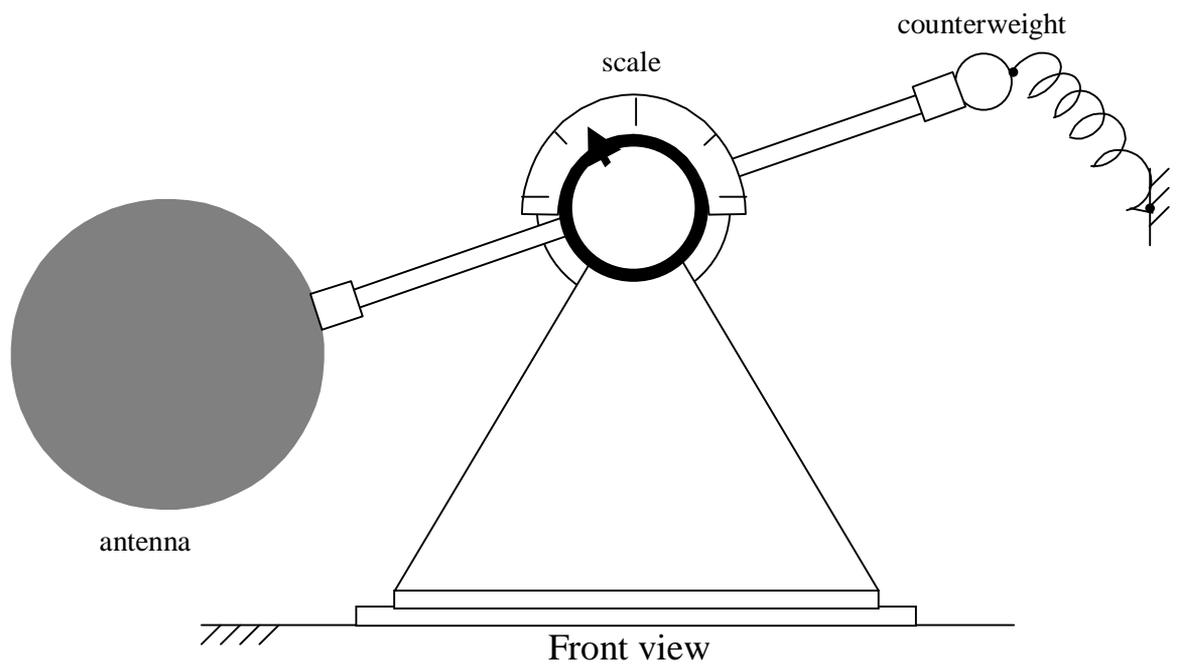

Front view

Fig.1 - Schematic diagram of the System H

| $I$ (A) | $m_{g(iron\ sphere)}$ (kg) | |
|---|---|---|
| | theory | experimental |
| 0.00 | 60.50 | 60.5 |
| 1.00 | 60.48 | 60.(4) |
| 2.00 | 60.27 | 60.(3) |
| 3.00 | 59.34 | 59.(4) |
| 4.00 | 56.87 | 56.(9) |
| 5.00 | 51.81 | 51.(9) |
| 6.00 | 43.09 | 43.(1) |
| 7.00 | 29.82 | 29.(8) |
| 8.00 | 11.46 | 11.(5) |
| 8.51 | 0.0 | 0.(0) |
| 9.00 | -12.16 | -12.(1) |
| 10.00 | -40.95 | -40.(9) |

Table 1

Note: The *inertial mass* of the iron sphere is $m_{iron\ sphere} = 60.50 kg$